# Electrical Relaxation and Transport in $0.5Cs_2O$-$0.5Li_2O$-$3B_2O_3$ Glasses

**Rahul Vaish** and **K. B. R. Varma**
Materials Research Centre
Indian Institute of Science
Bangalore 560 012, India

## ABSTRACT

The frequency and temperature dependence of the dielectric constant and the electrical conductivity of the transparent glasses in the composition $0.5Cs_2O$-$0.5Li_2O$-$3B_2O_3$ (CLBO) were investigated in the 100 Hz - 10 MHz frequency range. The dielectric constant for the as-quenched glass increased with increasing temperature, exhibiting anomalies in the vicinity of the glass transition and crystallization temperatures. The temperature coefficient of dielectric constant was estimated ($35 \pm 2$ ppm.$K^{-1}$) using Havinga's formula. The dielectric loss at 313 K is $0.005 \pm 0.0005$ at all the frequencies understudy. The activation energy associated with the electrical relaxation determined from the electric modulus spectra was found to be $1.73 \pm 0.05$ eV, close to that of the activation energy obtained for DC conductivity ($1.6 \pm 0.06$ eV). The frequency dependent electrical conductivity was analyzed using Jonscher's power law. The combination of these dielectric characteristics suggests that these are good candidates for electrical energy storage device applications.

## 1 INTRODUCTION

Glass-ceramics are generally obtained by the controlled crystallization of their corresponding glasses. Among this class of engineered materials, glasses comprising nano/micro-crystallites of polar phases have become increasingly important owing to their varied physical properties that include electrooptic, pyroelectric, piezoelectric and ferroelectric [1-3]. With their excellent electrical breakdown strength, these glass-ceramics are promising candidates for high energy density capacitor applications [4]. This is due to the fact that glass-ceramics have low/zero porosity and consequently high dielectric breakdown strength. Yet another important aspect is that these could be obtained with relative ease. They are also of technological prominence because of the flexibility that the glass-ceramic route offers in accomplishing the desired nano/microstructure by apposite heat-treatment. Indeed it is possible to tailor the glass-ceramics to exhibit the specific physical properties by strictly controlling the crystallite size, volume fraction of crystallization and their connectivity. A number of glasses comprising ferroelectric crystalline phases have been fabricated and investigated from their physical properties viewpoint [5-7].

Borate-based polar single crystals exhibit vital physical properties such as good optical transparency in the wide range of wavelengths, moderate melting temperature and high laser damage threshold. In particular, these crystals are promising for non-linear optical, pyroelectric piezoelectric and surface acoustic wave (SAW) device applications. Various borate-based single crystals such as $LiB_3O_5$, $Li_2B_4O_7$, $BaB_2O_4$, $BiB_3O_6$ were reported to possess promising physical properties [8-10]. $CsLiB_6O_{10}$ single crystals were demonstrated to exhibit excellent non-linear optical properties [11]. We have been making systematic attempts to fabricate glasses and glass-ceramics in the $Cs_2O$-$Li_2O$-$B_2O_3$ system and visualize their physical properties. The compositions were chosen such a way that are obtained $CsLiB_6O_{10}$ crystalline phase was obtained on heat-treatment at appropriate temperature. To begin with, the dielectric behavior of $0.5Cs_2O$-$0.5Li_2O$-$3B_2O_3$ (CLBO) glasses has been investigated, the details of which are reported in this article.

## 2 EXPERIMENTAL

Transparent glasses in the composition $0.5Cs_2O$-$0.5Li_2O$-$3B_2O_3$ (in molar ratio) were fabricated via the conventional melt-quenching technique. For this, $Cs_2CO_3$, $Li_2CO_3$ and $H_3BO_3$ were mixed and melted in a platinum crucible at 1323 K for 1h. The batch weight was 10 gm. Melts were quenched by pouring on a steel plate that was maintained at 423K and pressed with another plate to obtain 1-1.5 mm thick glass plates. All these samples were annealed at 623 K (6 h) which is well below the glass transition temperature. X-ray powder diffraction (XRD, Philips PW1050/37, Cu *K*α radiation) study was performed on the as-quenched powdered samples at room temperature to confirm their amorphous nature. The differential scanning calorimetry (DSC Model: Diamond DSC, Perkin-Elmer) runs were carried out in the 300 K –873 K temperature range. The glasses under study were heated at temperatures corresponding to their crystallization temperatures (as assessed by the DSC studies) for a few hours and analyzed by XRD studies.

The capacitance and dielectric loss (*D*) measurements on the as-quenched (annealed) polished glass plates that were silver painted were done using impedance gain phase analyzer (HP 4194 A) in the 100 Hz-10 MHz frequency range with a signal strength of 0.5 $V_{rms}$ at various temperatures (310–873 K). Thin silver leads were bonded to the sample using silver epoxy. Based on these data, the dielectric constant was evaluated by taking the dimensions and electrode geometry of the sample into account.

## 3 RESULTS AND DISCUSSION

The X-ray diffraction (Figure1) pattern that is obtained for the pulverized as-quenched sample confirms its overall amorphous nature. In Figure 2, the typical DSC trace recorded for CLBO glass-plates at a heating rate of 10 K/min is shown.

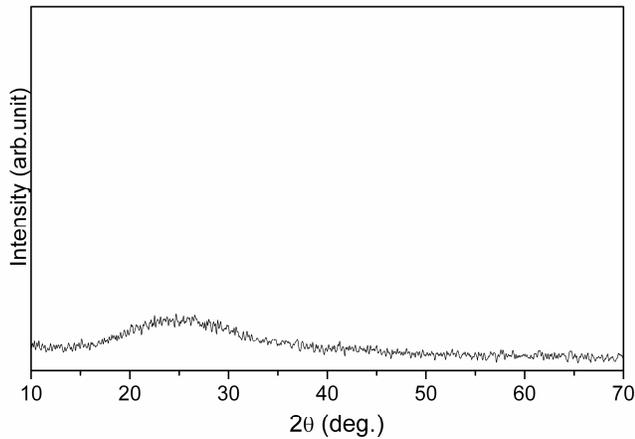

**Figure 1.** XRD pattern for the as-quenched pulverized CLBO glasses.

The endotherm around 746 K, followed by an exotherm at 807 K, are associated with the glass transition ($T_g$) and the onset of crystallization ($T_{cr}$) temperatures of the as-quenched samples, respectively. The variation of the dielectric constant ($\varepsilon_r^{'}$) with frequency (1 kHz-10 MHz) of measurement for CLBO glasses at different temperatures is shown in Figure 3a. The dielectric constant at 313 K, 373 K and 423 K is almost independent of frequency under study. At the other temperatures, $\varepsilon_r^{'}$ decreases with increase in frequency. The decrease is significant especially at low frequencies which may be associated with the mobile ion polarization combined with electrode polarization.

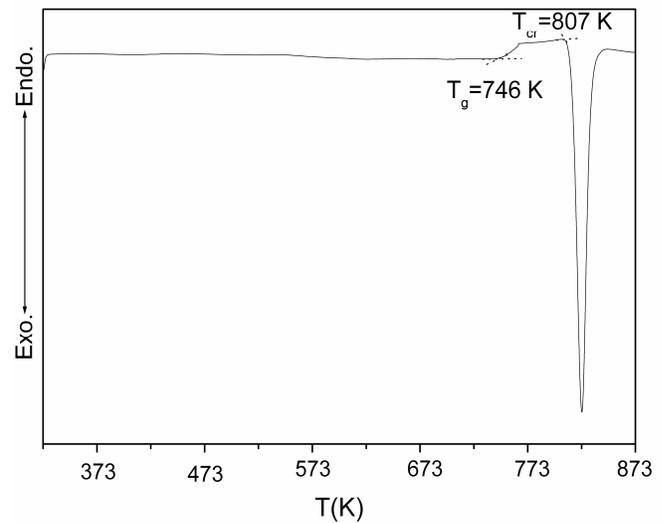

**Figure 2.** The DSC trace for CLBO glass-plate at a heating rate of 10 K /min.

The low frequency dispersion of $\varepsilon_r^{'}$ gradually increases with increase in temperature which may be due to an increase in the electrode polarization as well as thermal activation of $Li^+$ ions in the CLBO glasses. When the temperature rises, the dispersion region shifts towards higher frequencies and the nature of the dispersion changes at low frequencies due to the electrode polarization [Figure 3a]. As the frequency increases, $\varepsilon_r^{'}$ decreases due to high periodic reversal of the field, which reduces the contribution of the charge carriers towards the dielectric constant.

Figure 3b shows the variation of imaginary part of dielectric constant ($\varepsilon_r^{''}$) as a function of the frequency (1 kHz – 10 MHz) at various temperatures (313 – 623 K). The imaginary part of the dielectric constant corresponds to a current density within the dielectric that is no longer exactly π/2 out of phase with the electric field. It is responsible for the dissipation in the dielectric at the specific frequencies as depicted in Figure 3b. It is seen from the figure that the $\varepsilon_r^{''}$ decreases with the increase in frequency at all temperatures under study. It is known that the contribution to the dielectric loss generally consists of (a) conduction and (b) relaxation components. The imaginary part of dielectric constant can be explained as $\left[\sigma_{DC}/(\omega\varepsilon_o) + \varepsilon_{AC}^{''}\right]$ where $\varepsilon_0$ is the vacuum dielectric constant, $\omega$ (2π*f*) is the frequency and $\varepsilon_{AC}^{''}$ is the dielectric loss due to only relaxation process. The higher values of $\varepsilon_r^{''}$ at relatively low frequency may be attributed to the contribution arising from both the conduction and relaxation losses. At higher frequencies the relaxation losses are the only sources of dielectric loss. It is also noticed that the $\varepsilon_r^{''}$ increases with increase in temperature. As the temperature increases, the relaxation loss component reduces and the conduction loss component increases more rapidly. The dielectric loss (tan $\delta = \varepsilon_r^{''}/\varepsilon_r^{'}$) at 313 K is 0.005±0.0005 at all the frequencies under study.

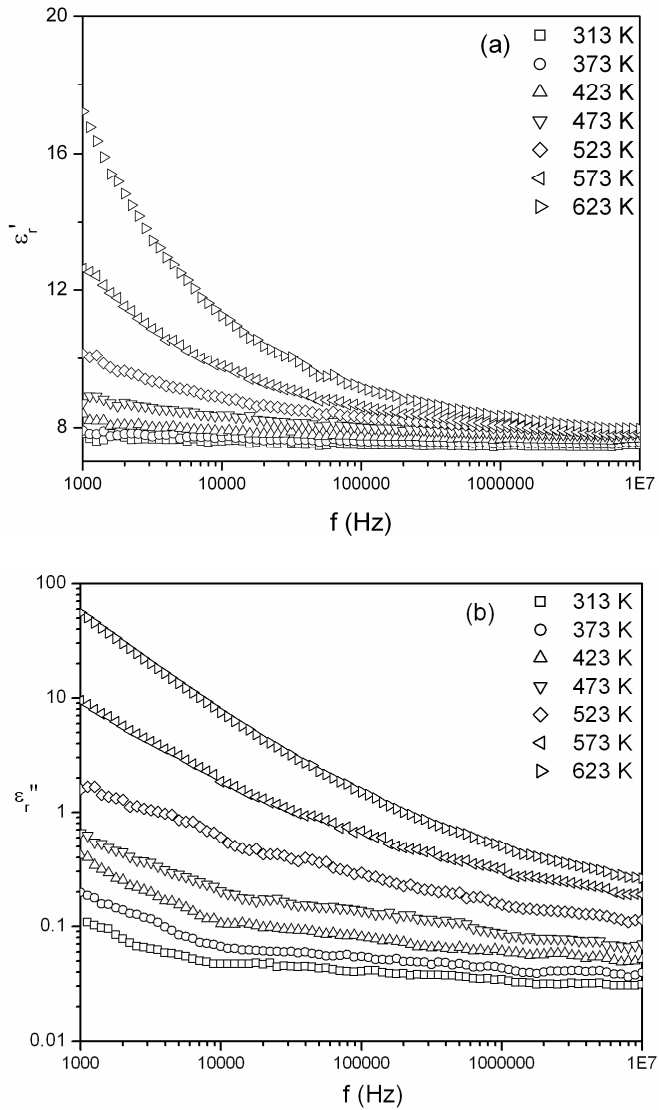

**Figure 3.** Frequency dependence of the (a) real component of dielectric constant and (b) imaginary component of dielectric constant of CLBO glasses at various temperatures.

The temperature dependence of dielectric constant for the as-quenched glasses was also investigated in the 310-450 K temperature range over the frequency range 100 kHz-10 MHz (Figure 4). It is noted from the figure that, at each frequency, the increment in dielectric constant with temperature (310 K-450 K) is approximately linear. The temperature dependence of the dielectric constant at different frequencies was rationalized using Havinga's model [12].

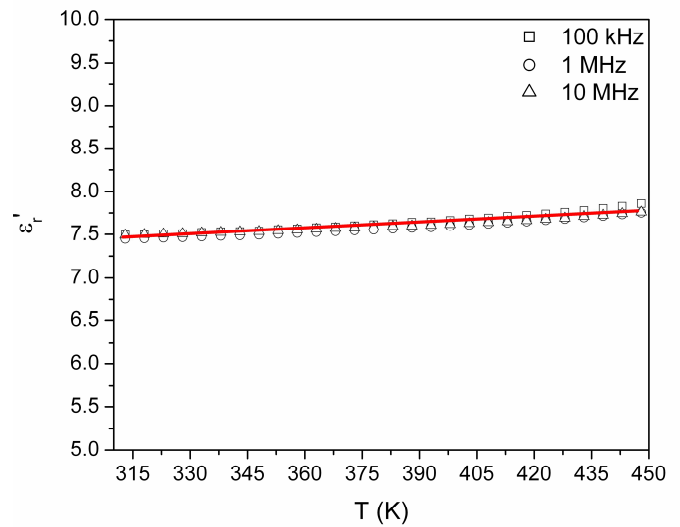

**Figure 4.** Variation of dielectric constant with temperature at various frequencies and solid line is a linear fit.

Havinga showed the temperature dependence of the dielectric constant at constant pressure, $P$ as follows;

$$\frac{1}{(\varepsilon_r^{'}-1)(\varepsilon_r^{'}+2)}\left(\frac{\partial \varepsilon_r^{'}}{\partial T}\right)_P = A + B + C \qquad (1)$$

According to this model, the temperature variation of the dielectric constant for the perfect solid dielectrics depends on three factors ($A$, $B$ & $C$) that are related to the thermal expansion and polarizability of the material. In the above relation "$A$" represents the decrease in the number of polarizable particles per unit volume as the temperature increases and has a direct effect on the volume expansion. "$B$" signifies the increase in the polarizability of a constant number of particles as the volume increases and "$C$" denotes the change in polarizability due to temperature changes at a constant volume.

In the present study, the temperature coefficient of dielectric constant $[(\varepsilon_r^{'}-1)(\varepsilon_r^{'}+2)]^{-1}(\partial \varepsilon_r^{'}/\partial T)_P$ for the linear region of dielectric constant (Figure 4) in the 310-450 K temperature range was estimated. The value of the coefficient is almost independent in the 100 kHz-10 MHz frequency and 310 -450 K temperature ranges. The average value is $35 \pm 2$ ppm.K$^{-1}$. This ensures temperature stability of these glasses upto 450K and can utilized for high temperature applications. Charaacteristics like low dielectric losses, high temperature stability along with zero piezoelectric noise are promising for capacitor applications. Inorder to have further insight into the dielectric behavior of CLBO glasses, its temperature dependent behavior at various frequencies is studied [Figures 5 a & 5b]. The dielectric constant is found to increase gradually in the 450-700 K temperature range as shown in the Figure 5a and subsequently increases rapidly above the glass transition at all the frequencies under study. This is due to the fact that above the glass transition temperature, the viscosity of the glass continuously decreases and facilitates easy response of the ions to the external electric field which results in a rapid increase in the dielectric constant. There is a decrease in the dielectric constant with the increase in frequency at all the temperatures under study. As the temperature increases, the dielectric constant further increases and exhibits a sharp peak around the crystallization temperature of CLBO glasses. The increase in the dielectric constant is attributed to an increase in the interfacial polarization. During the crystallization process the interfaces (that would get created) between the glassy regions and the crystallites having different dielectric constants and conductivities could be the origin of charge accumulation and interfacial polarization. The sharp rise in the dielectric constant might also due to the ions that indulge in rapid movement to transform from a random (glassy) to the ordered (crystalline) state associated with the conduction related polarization during the crystallization. The decrease in dielectric constant above 825 K is due to the reduction in the interfacial polarization and slow ionic movement as the glass is fully crystallized at this temperature.

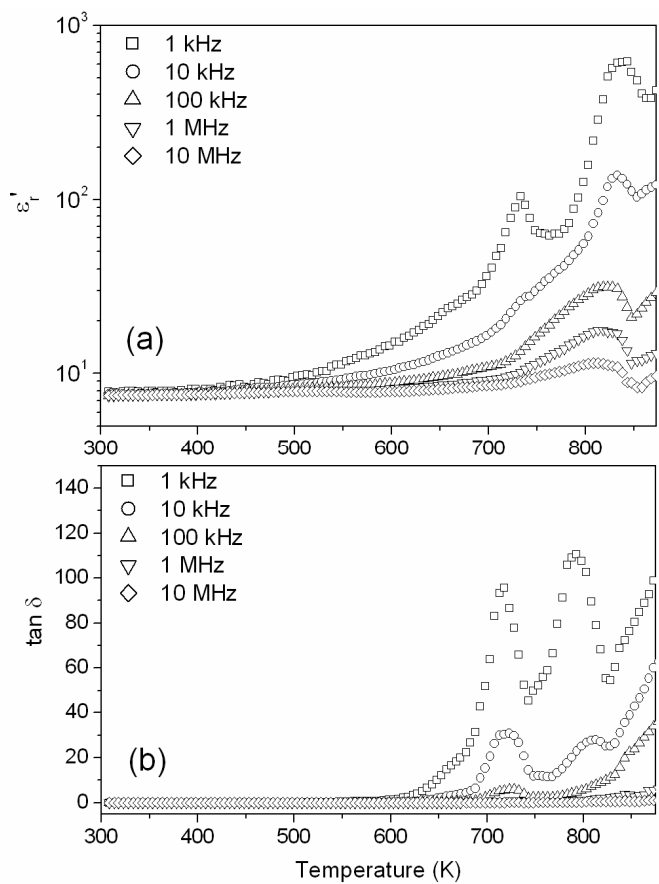

**Figure 5.** Temperature dependence of the (a) dielectric constant and (b) loss at various frequencies.

Subsequent jump in the dielectric constant above 850 K is related to $Li^+$ ion conduction-related polarization mechanism in the fully crystallized glass. It is essential to investigate into the behavior of the dielectric loss (tan $\delta$) with temperature. The dielectric loss versus temperature at various frequencies is depicted in Figure 5(b). It is found that tan $\delta$ increases with increasing temperature because of fast movement of the ions in the glass network and their increased response to an external electric field with increasing temperature. The tan $\delta$ abruptly increases at 640 K because of a sudden drop in the viscosity at glass transition. As the temperature increases from 750 K, tan $\delta$ increases due to an increase in conductivity of glass which arises from at least two different mechanisms. The first one is due to thermal activation in which the conductivity increases with increases in temperature according to the Arrhenius law. The second one originates from the structural changes that occur in the glass with temperature. During the crystallization process, one would expect more ionic diffusion within the glass matrix that enhances the electrical conductivity and consequently higher tan $\delta$. After completion of crystallization, tan $\delta$ decreases over a narrow range of temperatures due to the modifications that are likely to occur to the conducting paths within the sample. Subsequently, there is a rapid increase in tan $\delta$ with increase in temperature which may be due to the domination of AC conduction mechanism.

Electric modulus formalism was also invoked to rationalize the dielectric response of the present glasses. The use of electric modulus approach helps in understanding the bulk response of moderately conducting samples. This would facilitate to circumvent the problems caused by electrical conduction which might mask the dielectric relaxation processes. The complex electric modulus ($M^*$) is defined in terms of the complex dielectric constant ($\varepsilon^*$) and is represented as [13]:

$$M^* = (\varepsilon^*)^{-1} \qquad (2)$$

$$M^* = M' + iM'' = \frac{\varepsilon_r'}{(\varepsilon_r')^2 + (\varepsilon_r'')^2} + i\frac{\varepsilon_r''}{(\varepsilon_r')^2 + (\varepsilon_r'')^2} \qquad (3)$$

where $M'$, $M''$ and, $\varepsilon_r'$, $\varepsilon_r''$ are the real and imaginary parts of the electric modulus and dielectric constants, respectively. The real and imaginary parts of the modulus at different temperatures are calculated using Eq. 3 for the CLBO glasses and depicted in Figures 6a & 6b, respectively. One would notice from Figure 6a that at low frequencies, $M'$ approaches zero at all the temperatures under study suggesting the suppression of the electrode polarization. $M'$ reaches a maximum value corresponding to $M_\infty = (\varepsilon_\infty)^{-1}$ due to the relaxation process. It is also observed that the value of $M_\infty$ decreases with the increase in temperature. The imaginary part of the electric modulus (Figure 6b) is indicative of the energy loss under electric field. The $M''$ peak shifts to higher frequencies with increasing temperature. This evidently suggests the involvement of temperature dependent relaxation processes in the present glasses. The frequency region below the $M''$ peak indicates the range in which $Li^+$ ions drift to long distances. In the frequency range

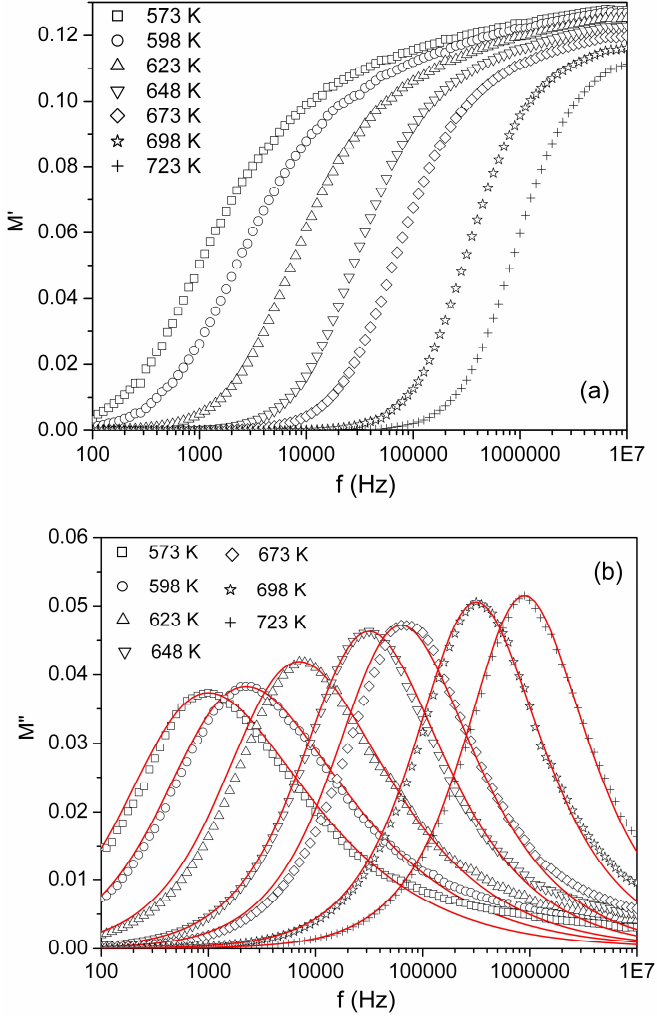

where $\tau_m$ is the conductivity relaxation time and the exponent $\beta$ (0 1) indicates the deviation from Debye-type relaxation. The value of $\beta$ could be determined by fitting the experimental data in the above equations. But it is desirable to reduce the number of adjustable parameters while fitting the experimental data. Keeping this point in view, the electric modulus behavior of the present glass system is rationalized by invoking modified KWW function suggested by Bergman. The imaginary part of the electric modulus ($M''$) is defined as [16]:

$$M'' = \frac{M''_{Max}}{(1-\beta) + \frac{\beta}{1+\beta}\left[\beta(\omega_{Max}/\omega) + (\omega/\omega_{Max})^\beta\right]} \quad (6)$$

where $M''_{Max}$ is the peak value of the $M''$ and $\omega_{Max}$ is the corresponding frequency. The above equation (Eq. 6) could effectively be described for $\beta \geq 0.4$. Theoretical fit of Eq. 6 to the experimental data is shown in Figure 6b as the solid lines. It is seen that the experimental data are well fitted to this model except in the high frequency regime. From the fitting of $M''$ versus frequency plots, the value of $\beta$ was determined and found to be temperature dependent. The plot of $\beta$ versus temperature is depicted in Figure 7. $\beta$ increases gradually with the increase in temperature indicating that as the temperature increases the glass network loosens and the interactions between $Li^+$ an $Cs^+$ ions and surrounding matrix decreases.

The relaxation time associated with the process was determined from the plot of $M''$ versus frequency. The activation energy involved in the relaxation process of ions could be obtained from the temperature dependent relaxation time as:

$$\tau_m = \tau_o \exp\left(\frac{E_R}{kT}\right) \quad (7)$$

where $E_R$ is the activation energy associated with the relaxation process, $\tau_o$ is the pre-exponential factor, $k$ is the Boltzmann constant and $T$ is the absolute temperature. Figure 8 shows a plot between $\ln(\tau_m)$ and $1000/T$ along with the theoretical fit (solid line) to the above equation (Eq. 7). The value that is obtained for $E_R$ is $1.73 \pm 0.05$ eV, which is ascribed to the motion of $Li^+$ and $Cs^+$ ions.

AC conductivity at different frequencies and temperatures, was determined by using the dielectric data using the following formula:

$$\sigma_{AC} = \omega \varepsilon_o D\varepsilon_r' \quad (8)$$

where $\sigma_{AC}$ is the AC conductivity at a frequency $\omega$ ($=2\pi f$). The frequency dependence of the AC conductivity at different temperatures is shown in Figure 9. At low frequency, the conductivity shows a flat response which corresponds to the DC part of the conductivity. At higher frequencies, the conductivity shows a dispersion. It is clear from the figure that the flat region increases with the increase in temperature. The phenomenon of the conductivity dispersion in solids is generally analyzed using Jonscher's law [17] which is above the peak, the ions are spatially confined to potential wells and free to move within the wells. The frequency range where the peak occurs is suggestive of the transition from long-range to short-range mobility. The electric modulus ($M^*$) could be expressed as the Fourier transform of a relaxation function $\phi(t)$:

$$M^* = M_\infty \left[1 - \int_0^\infty \exp(-\omega t)\left(-\frac{d\phi}{dt}\right)dt\right] \quad (4)$$

where the function $\phi(t)$ is the time evolution of the electric field within the materials and is usually taken as the Kohlrausch-Williams-Watts (KWW) function [14,15]:

$$\phi(t) = \exp\left[-\left(t/\tau_m\right)^\beta\right] \quad (5)$$

$$\sigma_{AC} = \sigma_{DC} + A\omega^n \quad (9)$$

where $\sigma_{DC}$ is the DC conductivity, $A$ is the temperature dependent constant and $n$ is the power law exponent which generally varies between 0 and 1. The exponent $n$ represents the degree of interaction between the mobile ions. The present glasses are found to obey the above mentioned universal power law at all the temperatures and frequencies under study. The theoretically fitted lines of Eq. 9 to the experimental data are shown in Figure 9 (solid lines). Eq. 9 is not fitted well in low frequency regime (Figure 9) due to electrode polarization.

**Figure 6.** (a) Real and (b) imaginary parts of the electric modulus as a function of frequency at different temperatures. The solid lines are the theoretical fits.

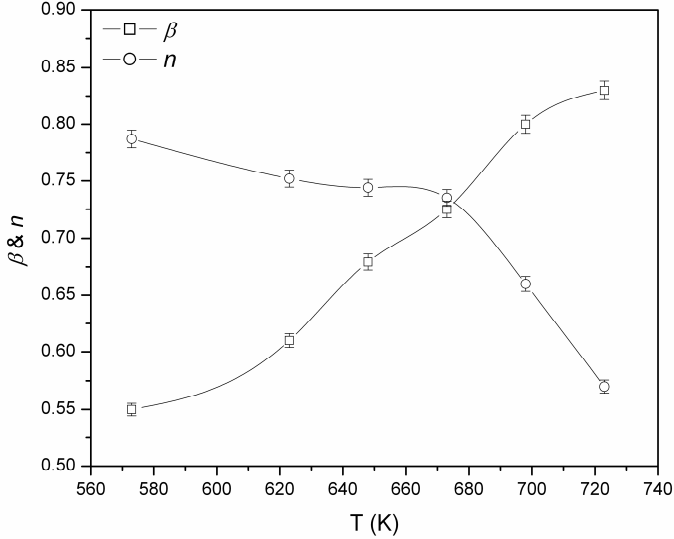

**Figure 7.** Variation of $\beta$ and $n$ with temperature

The variation of exponent $n$ as a function of temperature is depicted in Figure 7. It is known that the conductivity mechanism in any material could be understood from the temperature dependent behavior of $n$. To ascertain the electrical conduction mechanism in the materials, various models have been proposed [18]. These models include quantum mechanical tunneling model (QMT), the overlapping large-polaron tunneling model (OLPT) and the correlated barrier hopping model (CBH). According to the QMT model, the value of exponent $n$ is found to be 0.8 and increases slightly with increase in the temperature whereas the OLPT model predicts the frequency and temperature dependence of $n$. In the CBH model, the temperature dependent behavior of $n$ is proposed. This model states that the charge transport between localized states due to hopping over the potential barriers and predicts a decrease in the value of $n$ with the increase in temperature, which is consistent with the behavior of $n$ for the glasses under study (Figure 7).
This suggests that the conductivity behavior of CLBO glasses can be explained using correlated barrier hopping model. The present glasses do not seem to follow the Ngai's relation ($\beta = 1-n$) [19] as the plots of imaginary part of electric modulus are not fitted exactly in the high frequency regime which influences the value of $\beta$ [Figure 6 (b)]. Since the values for $\beta$ and $n$ are estimated in different frequency regions (as they could not be fitted well in the same frequency region), it is inconsistent with the Ngai's relation. Although the qualitative changes in the values of $\beta$ and $n$ are in conformity with the fact that both parameters represent the interaction between the ions.

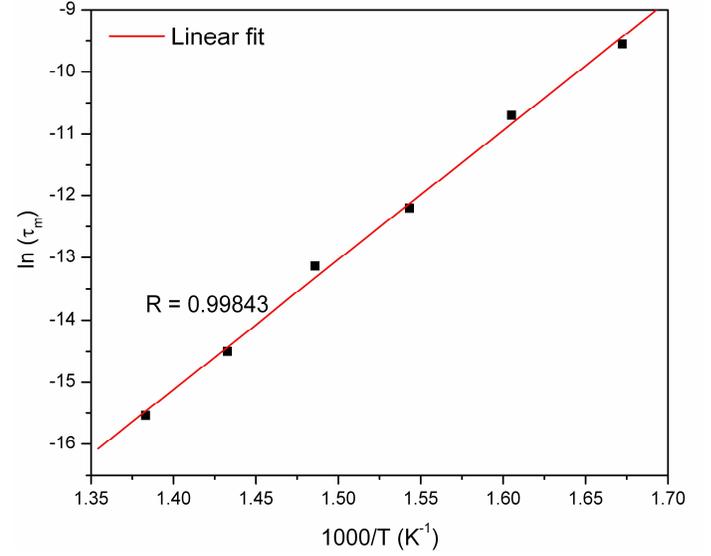

**Figure 8.** ln ($\tau_m$) versus 1000/T for CLBO glasses

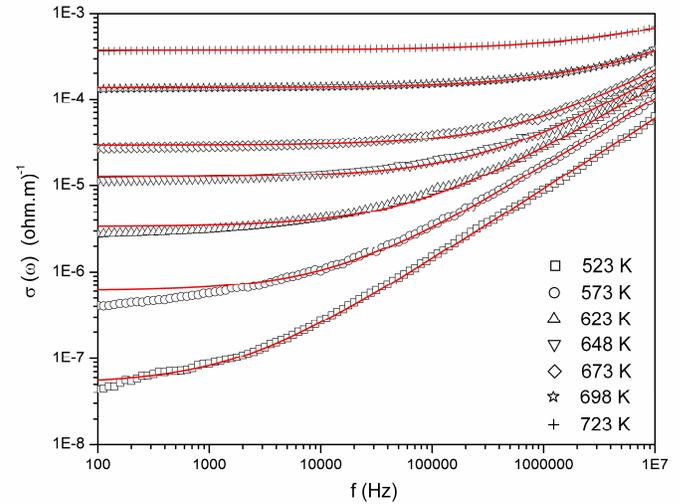

**Figure 9.** Variation of AC conductivity as a function of frequency at different temperatures and solid lines are the fitted curves.

Figure 10 shows the DC conductivity data obtained from the fitting (Eq. 9) of experimental data (Figure 9) at various temperatures. The activation energy for the DC conductivity was calculated from the plot of ln ($\sigma_{DC}$) versus 1000/T for the as-quenched CLBO glasses, which is shown in Figure 10. The plot is found to be linear and fitted using following Arrhenius equation,

$$\sigma_{DC}(T) = B \exp\left(-\frac{E_{DC}}{kT}\right) \quad (10)$$

where $B$ is the pre-exponential factor, $E_{DC}$ is the activation energy for the DC conduction. The activation energy for the as-quenched glass was calculated from the slope of the fitted line and found to be 1.6 ± 0.06 eV. The activation energies (DC conduction and relaxation) that are obtained unambiguously suggest that similar energy barriers are involved in both the relaxation and conduction processes.

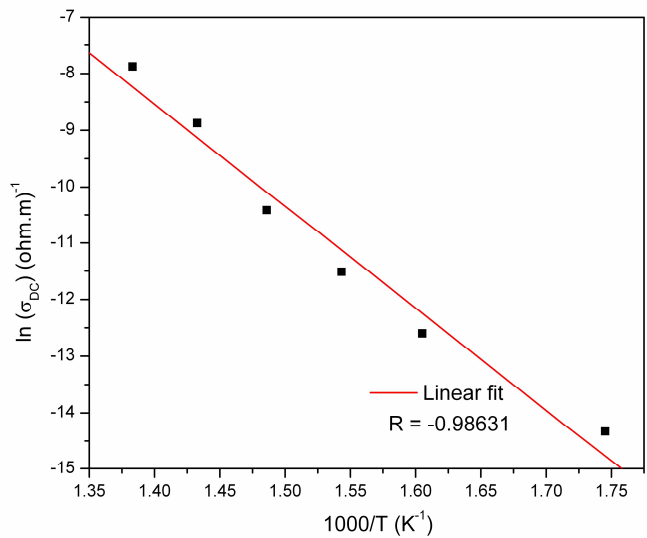

**Figure 10.** Arrhenius plot for DC conductivity

It is to be noted that the value of activation energy is higher than that of the activation energy obtained for $Li_2O-3B_2O_3$ glasses [20]. This is due to the fact that the $Li^+$ and $Cs^+$ ions in the CLBO glasses are randomly mixed in all the conduction pathways and each cation creates its own chemical environment [21]. These cation species prefer to migrate via pathways of sites adjusted to its own requirements. The ionic mobility is decreased if the pathways of the respective species interfere with each other. This blocking considerably reduces the conductivity of CLBO glasses in comparison to the corresponding single cation glasses ($Li_2O-3B_2O_3$). These effects result in higher activation energies associated with the conduction in CLBO glasses than that of the activation energy for $Li_2O-3B_2O_3$ glasses.

## 4 CONCLUSIONS

The activation energy associated with the conductivity relaxation time determined from the electric modulus spectra was found to be 1.73 ± 0.05 eV, close to that of the activation energy for DC conductivity (1.6 ± 0.06eV). It suggests that the movements of $Li^+$ and $Cs^+$ ions are responsible for both ionic conduction as well as the relaxation process. Temperature dependent behavior of the frequency exponent (*n*) suggested that the correlated barrier hopping model was the most appropriate to rationalize the electrical transport phenomenon in CLBO glasses. These glasses have fairly low temperature coefficient of dielectric (35ppm.$K^{-1}$) and reasonably low dielectric loss (0.005) at room temperature. These results suggest that the CLBO glasses are promising materials for capacitor applications.

## REFERENCES

[1] N. Shyam Prasad and K.B.R.Varma, "Dielectric, structural and ferroelectric properties of strontium borate glasses containing nanocrystalline bismuth vanadate", J. Mater. Chem., Vol. 11, pp. 1912-1918, 2001

[2] G. S. Murugan, K. B. R. Varma, Y. Takahashi and T. Komatsu, " Nonlinear-optic and ferroelectric behavior of lithium borate–strontium bismuth tantalate glass–ceramic composite", Appl. Phys. Lett., Vol 78, pp. 4019-402, 2001.

[3] G. S. Murugan and K. B. R. Varma,"Lithium borate– strontium bismuth tantalate glass nanocomposite: a novel material for nonlinear optic and ferroelectric applications", J. Mater. Chem., Vol. 12, pp. 1426-1436, 2002.

[4] N. J. Smith, B. Rangarajan, M. T. Lanagan and C. G. Pantano, "Alkali-free glass as a high density dielectric material", Mater. Lett., Vol. 63, pp.1245-1248, 2009.

[5] N. S. Prasad and K. B. R. Varma, "Nanocrystallization of $SrBi_2Nb_2O_9$ from glasses in the system $Li_2B_4O_7$-SrO-$Bi_2O_3$-$Nb_2O_5$", Mater. Sci.Engg. B, Vol. 90, pp. 246-253, 2002.

[6] N. S. Prasad, K. B. R. Varma and S. B. Lang, "Dielectric anomaly in strontium borate-bismuth vanadate glass nanocomposite", Phys. Chem. Solids, Vol. 62, pp. 1299-1311, 2001.

[7] C. Karthik and K. B. R. Varma," Evolution of nanocrystalline $BaBi_2Nb_2O_9$ in $Li_2B_4O_7$–BaO–$Bi_2O_3$–$Nb_2O_5$ glass system", J. Non-Cryst. Solids, Vol. 353, pp. 1307-1310, 2007.

[8] P. Becker, "Borate materials in nonlinear optics**"**, Adv. Mater., Vol. 10, pp. 979-992, 1998.

[9] D. N. Nikogosyan, "Lithium triborate (LBO)", Appl. Phys. A, Vol. 58, pp.181-190, 1994.

[10] I. Martynyuk-Lototska, T. Dudok, O. Mys and R. Vlok, "Elastic, piezooptic and acoustooptic properties of $SrB_4O_7$ and $PbB_4O_7$ crystals", Opt. Mater. Vol. 31, pp. 660-667, 2009.

[11] A. Majchrowski, I. V. Kityk and J. Kasperczyk, T.Ukasiewicz and A. Mefleh,"$CsLiB_6O_{10}$ crystallites embedded into the photopolymer matrices as promising materials for acoustically induced electrooptics", Mater.Lett., Vol. 50, pp.164-171, 2001.

[12] A. J. Bosman and E. E. Havinga, "Temperature dependence of dielectric

constants of cubic ionic compounds", Phys. Rev. Vol. 129, 1963, pp. 1593-1600, 1963.
[13] C. T. Moynihan, L. P. Boesch, and N. L. Laberge, ''Decay function for the electric field relaxation in vitreous ionic conductors'', Phys. Chem. Glasses, Vol. 14, pp. 122–125, 1973.
[14] G. Williams and D.C. Watts," Non-symmetrical dielectric relaxation Behaviour arising from a simple empirical decay function", Trans. Faraday Soc., Vol. 66, pp. 80-85, 1970.
[15] V. Provenzano, L. P. Boesch, V. Volterra, C. T. Moynihan and P. B. Macedo,"Electrical relaxation in $Na_2O \cdot 3SiO_2$ glass", J. Am. Ceram. Soc., Vol. 55, pp. 492-496, 1972.
[16] R. Bergman, "General susceptibility functions for relaxations in disordered systems", J. Appl. Phys., Vol. 88, pp. 1356-1365, 2000.
[17] A. K. Jonscher, "The universal dielectric response", Nature, Vol. 267, pp. 673- 679, 1977.
[18] A. Ghosh, "AC conduction in iron bismuthate glassy semiconductors", Phys. Rev. B, Vol. 42, pp.1388-1393, 1990.
[19] K. L. Ngai, J. N. Mundy, H. Jain, O. Kenert and G. Balzer-Jollenbeck, "Correlation between the activation enthalpy and Kohlrausch exponent for ionic conductivity in alkali aluminogermanate glasses", Phys. Rev. B.,Vol. 39, pp. 6169-6179, 1989.
[20] R. Vaish and K. B. R. Varma, "Dielectric properties of $Li_2O$-$3B_2O_3$ glasses", J. Appl.Phys., Vol. 106, pp. 064106, 2009.
[21] Y. Gao,"Mixed cation effect in $0.3[xLi_2O.(1-x)R_2O].0.7B_2O_3$ (R= Na, K, Rb) glasses", Chem. Phys. Lett., Vol. 417, pp. 430-433, 2006.